\documentclass[12pt,iopart,a4paper]{article}
\usepackage{amssymb}
\usepackage{amsfonts}
\usepackage{amsmath}
\usepackage{hyperref}
\usepackage[square,numbers,sort&compress,comma]{natbib}
\usepackage{graphicx}

\hypersetup{colorlinks=true, linkcolor=blue, citecolor=blue, urlcolor=blue}
\input{tcilatex}
\begin{document}

\author{P. G. C. Almeida, M. S. Benilov, and M. S. Bieniek \\
%EndAName
Departamento de F\'{\i}sica, Universidade da Madeira, \\
Largo do Munic\'{\i}pio, 9000 Funchal, Portugal,\smallskip\ and \\
Instituto de Plasmas and Fus\~{a}o Nuclear, IST, \ \\
Universidade de Lisboa, Portugal\\
}
\title{Modelling cathode spots in glow discharges in the cathode boundary
layer geometry }
\date{}
\maketitle

\begin{abstract}
Self-organized patterns of cathode spots in glow discharges are computed in
the cathode boundary layer geometry, which is the one employed in most of
the experiments reported in the literature. The model comprises conservation
and transport equations of electrons and a single ion species, written in
the drift-diffusion and local-field approximations, and Poisson's equation.
Multiple solutions existing for the same value of the discharge current and
describing modes with different configurations of cathode spots are computed
by means of a stationary solver. The computed solutions are compared to
their counterparts for plane-parallel electrodes, and experiments. All of
the computed spot patterns have been observed in the experiment. \newline
\end{abstract}

\section{Introduction}

\label{Introduction}

Self-organized patterns of cathode spots in DC glow microdischarges were
observed for the first time in 2004 [1] and represent an important and
interesting phenomenon. A range of experimental reports have since been
published [1--10]. Modelling of the phenomenon [9, 11--15] has revealed, in
agreement with the general theory of cathode spots and patterns in arc and
DC glow discharges [16], the existence of multiple steady-state solutions
for a given value of discharge current, which comprise modes of current
transfer associated with different cathode spot patterns. Predictions
concerning the existence of self-organizaed patterns in gases other than
xenon, generated from the modelling, have been confirmed experimentally [9].

The vast majority of the experiments [1--10] have been performed in a
discharge device comprising a flat cathode and a ring-shaped anode,
separated by a dielectric (cf., e.g., figure 1 of [10]); this discharge
configuration is called cathode boundary layer discharge (CBLD) by the
authors of the experiment. However, the modelling has been performed up to
date for discharges with parallel-plane electrodes only [9, 11--15]. The
question of how the shape of the anode affects the pattern of
self-organization has so far not been addressed. Furthermore, the effect
over 3D spots of absorption of the charged particles by a dielectric surface
has not been investigated in full due to computational difficulties [14].

In this work, 3D modelling of cathode spots is reported for the first time
in CBLD, and the full pattern of self-organization is computed with account
of absorption of charged particles at the dielectric surface. The outline of
the paper is as follows. The model is described in section \ref{Model and
Numerics}. In section \ref{Fundamental mode} the effect over the fundamental
mode of the discharge radius, the thickness of the cathode and dielectric,
and of a dielectric surface that reflects charged particles is investigated.
In section \ref{3dmodes} examples of computed 3D\ modes are given and
compared to their experimental counterparts. In section \ref{ref.walls} the
effect on 3D modes of a dielectric surface that reflects charged species is
investigated. In section \ref{Conclusions} conclusions are drawn.

\section{Model and Numerics}

\label{Model and Numerics}

The model employed in this work is the most basic self-consistent model of
glow discharge. Although the model is very well-known, it is described here
for the sake of completeness. The model comprises equations for conservation
of electrons and a single ion species, written in the drift-diffusion
transport approximation, and Poisson's equation.%
\begin{eqnarray}
\nabla \cdot \mathbf{J}_{i} &=&n_{e}\,\alpha \,\mu _{e}\,E-\beta
\,n_{e}\,n_{i},\;\;\;\mathbf{J}_{i}=-D_{i}\,\nabla n_{i}-n_{i}\,\mu
_{i}\,\nabla \varphi ,  \notag \\
\nabla \cdot \mathbf{J}_{e} &=&n_{e}\,\alpha \,\mu _{e}\,E-\beta
\,n_{e}\,n_{i},\;\;\;\mathbf{J}_{e}=-D_{e}\,\nabla n_{e}+n_{e}\,\mu
_{e}\,\nabla \varphi ,  \notag \\
\varepsilon _{0}\,\nabla ^{2}\varphi &=&-e\,\left( n_{i}-n_{e}\right) .
\label{1}
\end{eqnarray}%
\newline
Here $n_{i}$, $n_{e}$, $\mathbf{J}_{i}$, $\mathbf{J}_{e}$, $D_{i}$, $D_{e}$, 
$\mu _{i}$, and $\mu _{e}$ are number densities, densities of transport
fluxes, diffusion coefficients, and mobilities of the ions and electrons,
respectively; $\alpha $ is Townsend's ionization coefficient; $\beta $ is
coefficient of dissociative recombination; $\varphi $ is electrostatic
potential, $E=\left\vert \nabla \varphi \right\vert $ is electric field
strength; $\varepsilon _{0}$ is permittivity of free space; and $e$ is
elementary charge. The local-field approximation is employed (i.e.\ electron
transport and kinetic coefficients are assumed to depend on the local
electric field only).

Boundary conditions at the cathode and anode are written in the conventional
form. Diffusion fluxes of the attracted particles are neglected as compared
to drift; the normal flux of the electrons emitted by the cathode is related
to the flux of incident ions in terms of the effective secondary emission
coefficient $\gamma $, which is assumed to characterize all mechanisms of
electron emission (due to ion, photon, and excited atom bombardment) [17];
density of ions vanishes at the anode; electrostatic potentials of both
electrodes are given. The dielectric surface is electrically insulating, and
absorbs the charged particles (i.e. case\textit{\ i)} $n_{i}=n_{e}=0$); for
comparison, some solutions were computed for the case of a reflecting
dielectric surface (i.e. case \textit{ii)} $\ \mathbf{n\cdot }\nabla n_{i}=%
\mathbf{n\cdot }\nabla n_{e}=0$). With the computational domain from figure
1, the boundary conditions read\newline
\begin{eqnarray}
\text{cathode\ (}AB\text{)} &:&\text{ }\,\frac{\mathbf{\partial }n_{i}}{%
\partial z}=0,\text{\quad }J_{ez}=-\gamma J_{iz}\text{$,$\quad }\varphi =0%
\text{\thinspace };  \notag \\
\text{anode\ (}CDE\text{)} &:&\;\;\;n_{i}=0\text{$,$ }\frac{\partial n_{e}}{%
\partial n}=0\text{\thinspace },\text{\quad }\varphi =U;  \notag \\
\text{dielectric\ (}BC\text{)} &:&\;\;%
\begin{array}{l}
\text{\textit{i)}}\ n_{i}=n_{e}=0\ \, \\ 
\text{\textit{ii)} }\frac{\mathbf{\partial }n_{i}}{\partial r}=\frac{\mathbf{%
\partial }n_{e}}{\partial r}=0\,%
\end{array}%
,\;J_{er}-J_{ir}=0.  \label{2}
\end{eqnarray}%
Here $U$ is the discharge voltage, the subscripts $r$ and $z$ denote radial
and axial projections of corresponding vectors, and $\partial n_{e}/\partial
n$ is the normal derivative of electron density. The lengths $DE$ and $AG$
are large enough so that boundaries $EF$ and $FG$ do not affect the
solution. The control parameter can be either discharge voltage $U$ or
discharge current $I$, depending on the slope of the current
voltage-characteristics (CVC) $U\left( I\right) $.

Results reported in this work refer to a discharge in xenon under the
pressure of $30\,\mathrm{Torr}$. The (only) ionic species considered is $%
\mathrm{Xe}_{2}^{+}$. The transport and kinetic coefficients are the same as
in [14]. A more detailed model (one that also took into account both atomic
and molecular ions, excited atoms, excimers, stepwise ionization, ionization
of excimers and non-locality of electron energy) resulted in qualitatively
similar self-organized patterns [14], therefore the relatively simple model
described in this section was seen as sufficient for the purposes of this
investigation.

The problem (\ref{1}) to (\ref{2}) admits multiple solutions describing
different discharge modes. One such mode exists for all ranges of current,
it is 2D (axially symmetric) and termed fundamental, this is routine to
calculate. 3D modes bifurcate from (and rejoin) the fundamental mode and are
termed non-fundamental modes.

To calculate non-fundamental modes, one first locates points of bifurcation
on the fundamental mode by means of linear stability analysis, the procedure
is discussed in detail in [11]. Next a 3D calculation domain is created by
rotating the 2D axially symmetric domain from figure 1 about the axis of
symmetry, by an amount corresponding to the expected azimuthal periodicity
of the 3D mode being sought. The beginning of the non-fundamental mode is
then searched for on the fundamental mode, with the 3D\ calculation domain,
in the vicinity of the bifurcation point predicted by the linear stability
analysis. Small perturbations are introduced at the bifurcation point; the
iterations will eventually converge to the 3D mode. The remainder of the 3D
mode is straight-forward to calculate.

The above procedure was realized using stationary and eigenvalue solvers
from the commercial product COMSOL Multiphysics. The time taken by the
stationary solver to find convergence to one of the most computationally
intensive 3D solutions is around $1$ hour, with a computer with a Intel Core
i7-4770 CPU at $3.4\unit{GHz}$ and $32\,\mathrm{GB}$ of RAM.

\section{Results}

\label{Results}

\subsection{Fundamental mode}

\label{Fundamental mode}

In figure 2, the CVC of the fundamental mode is displayed in four sets of
conditions, labeled $1$ to $4$ in the figure. Surprisingly, two turning
points and a loop are present on the CVC corresponfing to the baseline
conditions, line 1.

The whole current range in figure 2 can be divided into three regions,
marked I, II, III. At the top of the figure there is an illustration of the
characteristic distribution of current density on the cathode surface for
each region. The color range shown in the bar is used also for the rest of
the document. The general pattern of evolution of the fundamental mode with
increasing current is as follows. In region I, corresponding to the Townsend
discharge, the current is distributed on the cathode in the form of a ring.
In region II, the ring of current grows thicker with increasing current. In
region III, corresponding to the abnormal discharge, the discharge fills
most of the cathode surface.

In the case represented by line 1, a pattern with a central spot appears on
the section between the turning points, as indicated in the figure. This
transition is accompanied by a loop in the CVC. The loop is absent in the
CVC of cases $2$ and $3$; the larger radius and the reflecting dielectric
surface, respectively, prevent the transition from a ring to central spot.
The loop is also absent in the CVC of case $4$. The CVC of cases $1$, $2$
and $4$ (the ones with absorbing dielectric surface) have small humps in
range I, although this cannot be seen in the scale of figure 2.

\subsection{3D modes}

\label{3dmodes}

Figure 3 displays the CVC of the fundamental mode for condition set $1$,
points of bifurcation of 3D modes, and an example 3D mode. Each pair of
points $a_{i}$ and $b_{i}$ designates from where a 3D\ modes branches off
from and rejoins the fundamental mode. Mode $a_{i}b_{i}$ possesses period $%
2\pi /i$, meaning that $a_{1}b_{1}$ possesses azimuthal period $2\pi $, mode 
$a_{2}b_{2}$ possesses azimuthal period $\pi $, and so on. Bifurcation
points $b_{1}$ to $b_{4}$ virtually coincide. Points $b_{5}$ and $b_{6}$ are
positioned on the section between the turning points. The 3D modes branch
off and rejoin the fundamental mode in a palindromic order along current,
which conforms to previous modelling of discharges with parallel-plane
electrodes.

As an example, the CVC of mode $a_{3}b_{3}$ is shown in figure 3. (The
schematic in the figure illustrates the pattern of spots associated with
this mode.) The CVC manifests a plateau between $60\unit{A}\unit{m}^{-2}$
and $300\unit{A}\unit{m}^{-2}$, which is a manifestation of the normal
current density effect. Note that the plateau also is present in the
computed mode of the same azimuthal period for a vessel with parallel-plane
electrodes and reflecting dielectric surface [15].

CVCs of different modes would be difficult to distinguish in figure 3. A
more convenient representation is shown in figure 4: the fundamental and
four non-fundamental modes are mapped in the plane $(\left\langle
j\right\rangle ,j_{c})$, where $j_{c}$ is current density at a point which
is positioned on the upper vertical radius (with respect to the orientation
of every schematic) at $r=0.4$ $\unit{mm}$, which is the radius of the ring
spot in the Townsend discharge. Note that such choice ensures maximum
distinction between the modes. Following the fundamental mode from low to
high currents, it is seen that $j_{c}$ decreases while the central spot is
forming, then it increases as the ring mode forms, thus yielding a limp
Z-shape on the bifurcation diagram. Modes $a_{5}b_{5}$, $a_{6}b_{6}$ possess
turning points.

In figure 5 the evolution is shown of spot patterns associated with modes $%
a_{3}b_{3},$\ $a_{4}b_{4},$ $a_{5}b_{5},$\ $a_{6}b_{6}$\ from figure 4 as
discharge current is changed. Let us consider first the evolution of the
patterns for mode $a_{3}b_{3}$ which is shown in figure 5a). The state $%
303.06\unit{A}\unit{m}^{-2}$ is positioned in the vicinity of the
bifurcation point $b_{3}$, the pattern is of three diffuse elongated spots,
slightly deforming into a 3D structure with the period of $2\pi /3$. At $%
296.06\unit{A}\unit{m}^{-2}$ the three spots have moved away from each other
and become more distinct, intense and bean-shaped. The spots then become
circular, and move farther from the center of the cathode as seen in state $%
36.1\unit{A}\unit{m}^{-2}$. The spots then once more become bean-shaped,
then once again gain a triangular type structure as in state $4.4\unit{A}%
\unit{m}^{-2}$. At state $4.4\unit{A}\unit{m}^{-2}\ $a central
triangle-shaped `cold spot' is present. Continuing along the mode with
decreasing current, the triangle shaped region becomes less sharp, and the
whole pattern becomes more like the ring-shaped distribution present at $%
a_{3}$.

The evolution of the patterns associated with modes $a_{4}b_{4}$, $%
a_{5}b_{5} $, and $a_{6}b_{6}$, is shown in figures 5b-d, respectively,
follows the same trend as mode $a_{3}b_{3}$: first the ring is transformed
into elongated bean-shaped spots and then circular spots, then they migrate
to a different radius, and there, from circular spots they turn into
bean-shaped spots and then merge into a different ring. No 3D modes with
central hot spots were found in the present work, while in previous
modelling they were; e.g.\ [15]. The images in figure 5 can be compared to
experimentally observed patterns of spots, figure 2 of [10]. The computed
evolution from the abnormal mode into mode $a_{4}b_{4}$, comprising four
spots, is in good agreement with the experimentally observed transition
between the abnormal mode into a mode comprising four spots.

In figure 2 of [10], it can be seen how the modes appear in the experiment:
starting from the abnormal mode and reducing discharge current, a mode
comprising four spots appears. As current is further reduced, modes
comprising five and six spots appear. Further reducing current from the mode
with six spots, the discharge goes back to modes with five, four, three and
a ring spot. In the modelling, cf.\ figure 3, starting from a state in the
abnormal mode and following the fundamental mode in the direction of low
currents, the bifurcation point $b_{1}$ of mode $a_{1}b_{1}$, comprising one
spot, appears first. The next bifurcation point to appear is $b_{2}$ of mode 
$a_{2}b_{2}$, comprising two spots; and so on until bifurcation point $b_{6}$
of mode $a_{6}b_{6}$, comprising six spots, following the same trend
observed in the experiment. On further following the fundamental mode in the
direction of low currents, eventually the bifurcation point $a_{5}$ of mode $%
a_{5}b_{5}$ appears; and so on until $a_{1}$ of mode $a_{1}b_{1}$, again
following the same trend as in the experiment.\emph{\ }

\subsection{Effect of a reflective dielectric surface}

\label{ref.walls}

In figure 6 mode\ 3 from figure 2 and three non-fundamental modes computed
with a dielectric surface that reflects ions and electrons are shown. (In
figure 6, $j_{c}$ is the current density occurring on the periphery of the
cathode at $r=R$). Similarly to the case of absorbing dielectric surface,
the patterns comprise the axially symmetric ring spot or circular
arrangements of 3D spots.

Both the axially symmetric ring spot and 3D spots are centered at the
periphery of the cathode in this case, similarly to what happens in the
similar modes in the parallel-plane electrode configuration with reflecting
lateral wall [15]. On the other hand, this is in contrast to what is
observed in the experiment and in the modelling with the dielectric surface
absorbing the charged particles, where the spots are shifted away from the
periphery of the cathode rather than centered at the periphery.

\section{Conclusions}

\label{Conclusions}

Self-organized 3D spot modes have been computed for cathode boundary layer
discharges. The general form of the computed self-organized patterns is
similar to those computed previously in the parallel-plane configuration and
to those observed in the experiment in the sense that all of them comprise
axially symmetric ring spots or circular arrangements of 3D spots. This is
consistent with experimental evidence [4] that similar self-organized
patterns appear in both electrode configurations.

Simulations of 3D spot patterns with the dielectric surface fully absorbing
the charged particles reveal spots not centered at the periphery of the
cathode, but rather located inside the cathode, as they are in the
experiment. It has been found that there is a palindromic series of the
number of spots with discharge current, which is consistent with
observations of switching between modes with different patterns in the
experiment [10].

\section{Acknowledgments}

The work was supported by FCT - Funda\c{c}\~{a}o para a Ci\^{e}ncia e a
Tecnologia of Portugal through the projects PTDC/FIS-PLA/2708/2012 and
Pest-OE/UID/FIS/50010/2013.

\bibliographystyle{apsrev4-1}
\bibliography{arc,benilov}

\section{References}

\ \ \ \ [1] K. H. Schoenbach, M. Moselhy, and W. Shi, Plasma Sources Sci.
Technol. 13, 177 (2004).

[2] M. Moselhy and K. H. Schoenbach, J. Appl. Phys. 95, 1642 (2004).

[3] N. Takano and K. H. Schoenbach, Plasma Sources Sci. Technol. 15, S109
(2006).

[4] N. Takano and K. H. Schoenbach, in Abstracts of the 2006 IEEE
International Con-ference on Plasma Science (IEEE, Traverse City, MI, USA,
2006) p. 247.

[5] B.-J. Lee, H. Rahaman, K. Frank, L. Mares, and D.-L. Biborosch, in Proc.
28th ICPIG (Prague, July 2007), edited by J. Schmidt, M. \v{S}imek, S. Pek%
\'{a}rek, and V. Prukner (Institute of Plasma Physics AS CR, ISBN
978-80-87026-01-4, Prague, 2007) pp. 900--902.

[6] W. Zhu, N. Takano, K. H. Schoenbach, D. Guru, J. McLaren, J. Heberlein,
R. May, and J. R. Cooper, J. Phys. D: Appl. Phys. 40, 3896 (2007).

[7] B.-J. Lee, D.-L. Biborosch, K. Frank, and L. Mares, J. Optoelectron.
Adv. Mater.10, 1972 (2008).

[8] K. H. Schoenbach and W. Zhu, IEEE J. Quantum. Electron. 48, 768 (2012).

[9] W. Zhu, P. Niraula, P. G. C. Almeida, M. S. Benilov, and D. F. N.
Santos, Plasma Sources Sci. Technol. 23, 054012 (2014).

[10] W. Zhu and P. Niraula, Plasma Sources Sci. Technol. 23, 054011 (2014).

[11] P. G. C. Almeida, M. S. Benilov, M. D. Cunha, and M. J. Faria, J. Phys.
D: Appl. Phys. 42, 194010 (21pp) (2009).

[12] P. G. C. Almeida, M. S. Benilov, and M. J. Faria, Plasma Sources Sci.
Technol. 19, 025019 (13pp) (2010).

[13] P. G. C. Almeida, M. S. Benilov, and M. J. Faria, IEEE Trans. Plasma
Sci. 39, 2190 (2011).

[14] P. G. C. Almeida and M. S. Benilov, Phys. Plasmas 20, 101613 (2013).

[15] P. G. C. Almeida, M. S. Benilov, and D. F. N. Santos, ArXiv e-prints
(2014), arXiv:1406.4394 .

[16] M. S. Benilov, Plasma Sources Sci. Technol. 23, 054019 (2014).

[17] Yu. P. Raizer, Gas Discharge Physics (Springer, Berlin, 1991).

\bigskip

\section{Figures}

Figure 1: Configuration of a cathode boundary layer vessel. AG is an axis of
symmetry.

Figure 2: Fundamental mode, Xenon, 30 Torr, CBLD configuration of figure 1.
1: h=0.5 mm, h$_{a}$=0.1 mm, R=0.5 mm, absorbing lateral wall. 2: h=0.5 mm, h%
$_{a}$=0.1 mm, R=1.5 mm, absorbing lateral wall. 3: h=0.5 mm, h$_{a}$=0.1
mm, R=0.5 mm, reflecting lateral wall. 4: h=0.25 mm, h$_{a}$=0.25 mm,
R=0.375 mm, absorbing lateral wall.

Figure 3: CVC, xenon, 30 Torr, h=0.5 mm, h$_{a}$=0.1 mm, R=0.5 mm, absorbing
lateral wall. Solid: Fundamental mode (mode 1 of figure 2). Dashed: mode a$%
_{3}$b$_{3}$. Circles: points of bifurcation.

Figure 4: Bifurcation diagram, xenon, 30 Torr, h=0.5 mm, h$_{a}$=0.1 mm,
R=0.5 mm, absorbing lateral wall. Solid: Fundamental mode (mode 1 of figure
2). Dashed: modes a$_{3}$b$_{3}$, a$_{4}$b$_{4}$, a$_{5}$b$_{5}$. Dotted:
mode a$_{6}$b$_{6}$. Circles: points of bifurcation.

Figure 5: Evolution of patterns associated with 3D modes of figure 4. a)
mode a$_{3}$b$_{3}$ b) mode a$_{4}$b$_{4}$ c) mode a$_{5}$b$_{5}$ d) mode a$%
_{6}$b$_{6}$.

Figure 6: Bifurcation diagram, xenon, 30 Torr, h=0.5 mm, h$_{a}$=0.1 mm,
R=0.5 mm, reflecting lateral wall. Solid: fundamental mode (mode 3 from
figure 2). Dashed: modes a$_{3}$b$_{3}$, a$_{6}$b$_{6}$, a$_{10}$b$_{10}$.
Circles: points of bifurcation.

\end{document}